\documentclass[12pt,preprint]{aastex}

\received{2005 December 15}
\begin{document}

\title{A Spectroscopic Survey of Faint Quasars in the SDSS Deep Stripe: I. 
  Preliminary Results from the Co-added Catalog\footnote{Observations 
  reported here were obtained at the MMT Observatory, a joint facility of 
  the Smithsonian Institution and the University of Arizona.}}

\author{Linhua Jiang\altaffilmark{1}, Xiaohui Fan\altaffilmark{1},
  Richard J. Cool\altaffilmark{1}, Daniel J. Eisenstein\altaffilmark{1}, 
  Idit Zehavi\altaffilmark{1},
  Gordon T. Richards\altaffilmark{2,3}, Ryan Scranton\altaffilmark{4}, 
  David Johnston\altaffilmark{2}, Michael A. Strauss\altaffilmark{2}, 
  Donald P. Schneider\altaffilmark{5}, and J. Brinkmann\altaffilmark{6}}
\altaffiltext{1}{Steward Observatory, University of Arizona,
  933 North Cherry Avenue, Tucson, AZ 85721}
\altaffiltext{2}{Department of Astrophysical Sciences, Peyton Hall, Princeton,
  NJ 08544}
\altaffiltext{3}{Department of Physics and Astronomy, The Johns Hopkins
  University, 3400 North Charles Street, Baltimore, MD 21218}
\altaffiltext{4}{University of Pittsburgh, Department of Physics and Astronomy,
  3941 O'Hara Street, Pittsburgh, PA 15260}
\altaffiltext{5}{Department of Astronomy and Astrophysics, Pennsylvania
  State University, 525 Davey Laboratory, University Park, PA 16802}
\altaffiltext{6}{Apache Point Observatory, P.O. Box 59, Sunspot, NM 88349}

\begin{abstract}

In this paper we present the first results of a deep spectroscopic 
survey of faint quasars in the Sloan Digital Sky Survey (SDSS) Southern
Survey, a deep survey carried out by repeatedly imaging a 270 deg$^2$ area.
Quasar candidates were selected from the deep data with good completeness
over $0<z<5$, and 2 to 3 magnitudes fainter than the SDSS main survey.
Spectroscopic follow-up was carried out on the 6.5m MMT with Hectospec. 
The preliminary sample of this SDSS faint quasar survey (hereafter SFQS) 
covers $\sim3.9$ deg$^2$, contains 414 quasars, and reaches $g=22.5$. 
The overall selection efficiency is $\sim$ 66\% ($\sim$ 80\% at $g<21.5$);
the efficiency in the most difficult redshift range ($2<z<3$) is better 
than 40\%. We use the $1/V_{a}$ method to derive a binned
estimate of the quasar luminosity function (QLF) and model the QLF using 
maximum likelihood analysis. The best model fits confirm previous results
showing that the QLF has 
steep slopes at the bright end and much flatter slopes 
($-1.25$ at $z\la 2.0$ and $-1.55$ at $z\ga 2.0$) at the faint end, 
indicating a break in the QLF slope. Using a 
luminosity-dependent density evolution model, we find that the quasar
density at $M_{g}<-22.5$ peaks at $z\sim 2$, which is later in
cosmic time than the peak of $z\sim 2.5$ found from surveys of more
luminous objects.
The SFQS QLF is consistent with the results of the 2dF QSO Redshift 
Survey, the SDSS, and the 2dF-SDSS LRG and QSO Survey, but probes fainter 
quasars. We plan to obtain more quasars from future observations and 
establish a complete faint quasar sample with more than 1000 objects over 
10 deg$^2$.

\end{abstract}

\keywords{galaxies: luminosity function --- quasars: general --- surveys}

\section{Introduction}

One of the most important properties of quasars is their strong evolution with
cosmic time. The quasar luminosity function (QLF) is thus of particular 
importance in understanding quasar formation and evolution and exploring 
physical models of quasars. It has been shown that quasar activity and the 
formation processes of galaxies and supermassive black holes (SMBHs) are 
closely correlated \citep[e.g.][]{kau00,wyi03,hop05,cro06}, so the QLF is 
essential to study galaxy formation, the accretion history of SMBHs during 
the active quasar phase, and its relation to galaxy evolution.
Quasars are strong X-ray sources, thus the QLF can provide important
constraints on the quasar contribution to the X-ray and ultraviolet background
radiation \citep[e.g.][]{koo88,boy98,mus00,wor05}. The QLF is also useful for
understanding the spatial clustering of quasars and its relation to quasar
life times \citep[e.g.][]{mar01}.

The differential QLF is defined as the density of quasars per unit comoving 
volume and unit luminosity interval as a function
of luminosity and redshift. If the redshift and luminosity dependence is 
separable, the QLF can be modeled in terms of pure density
evolution (PDE), pure luminosity evolution (PLE), or a combination of the two
forms. Earlier work found that there was a strong decline of quasar
activity from $z\sim2$ to the present universe, and a model with a single 
power-law shape provided good fits to the observed QLF for $z<2$ at the 
bright end \citep{mar83,mar84,mar85}. When more faint quasars were discovered,
a break was found in the luminosity function \citep[e.g.][]{koo88,boy88},
and the shape of QLF was modeled by a double power-law form with a steep
bright end and a much flatter faint end. In this double power-law model,
luminosities evolve as PLE, and density evolution is not necessary 
\citep[e.g.][]{boy88}. Some studies cast doubt on this claim, however. First,
the existence of the break in the QLF slope is not obvious
\citep[e.g.][]{haw95,gol98,wis00,wol03}. Second, it was found that the 
PLE model was not sufficient to describe the quasar evolution at $z<2$
\citep[e.g.][]{hew93,laf97,gol98,wis00}.

The density of luminous quasars reaches a maximum at $2<z<3$ (hereafter 
referred to as the mid-$z$ range), and drops rapidly toward higher redshift
\citep[e.g.][]{pei95}. The high-$z$ ($z>3$) QLF has been explored only for 
bright quasars \citep[e.g.][]{war94,ken95,sch95,fan01b}. At $z\sim4$, the 
QLF was fitted by a single power-law form with an exponential decline in
density with redshift \citep{sch95,fan01b}, and its slope is
much flatter than the bright-end slope of the QLF at $z<3$. This
indicates that, at high redshift, the shape of the QLF evolves with redshift
as well \citep{fan01b}. However, a large sample of high-$z$ quasars is needed
to prove this claim.

One of the largest homogeneous samples of low-$z$ quasars comes from the 2dF 
QSO Redshift Survey and 6dF QSO Redshift Survey 
\citep[hereafter 2QZ;][]{boy00,cro04}. The 2QZ
survey includes 25,000 quasars, and covers a redshift range of $0.4<z<2.1$
and a magnitude range of $16<b_{J}<20.85$. The QLF derived from the 2QZ
can be well fitted by a double power-law form with a steep slope at the bright
end and a flatter slope at the faint end, showing a break in the QLF. The
evolution is well described by the PLE. The Sloan Digital Sky Survey 
\citep[SDSS;][]{yor00} is collecting the largest spectroscopic samples of
galaxies and quasars to date. The SDSS main survey covers a large redshift
range from $z=0$ to $z>5$, however, it only selects bright objects: it targets 
quasars with $i<19.1$ at $z\la 3$ and $i<20.2$ at $z\ga 3$ \citep{ric02}. 
The QLF derived from SDSS Data Release Three \citep[DR3;][]{sch05}
shows some curvature at the faint end, but the survey does not probe faint 
enough to test the existence of the break \citep{ric06}. The best fit model 
of the SDSS-DR3 QLF shows, for the first time in a single large redshift 
range, the flattening of the QLF slope with increasing redshift.
Selecting $g<21.85$ and $z<2.5$ quasar candidates from SDSS imaging and 
spectroscopically observing them with the 2dF instrument,
the 2dF-SDSS LRG and QSO Survey (2SLAQ) probes deeper than 
either the SDSS or the 2QZ \citep{ric05}.
The QLF of the 2SLAQ is consistent with the 2QZ QLF from \citet{boy00}, but 
has a steeper faint-end slope than that from \citet{cro04}.

Despite the investigations described above, the optical QLF over both a 
large redshift range and a large luminosity range is far from well 
established. First, the faint-end slope of the QLF is still uncertain. Most 
wide-field surveys, including the SDSS, are shallow and can only sample the 
luminous quasars. Although the 2SLAQ survey probed to $g=21.85$, it only 
covers the low-$z$ range ($z\la 2.2$). Furthermore, the existence of the 
break in the QLF slope is uncertain, and different surveys give different 
slopes at the faint end \citep[e.g.][]{wol03,cro04,ric05}. Second, it is 
unclear whether PLE is sufficient to describe quasar evolution at low redshift.
Third, the high-$z$ ($z>3$) QLF has not been well established, especially at
the faint end. In addition, the density of luminous quasars peaks between
$z=2$ and $z=3$, yet the colors of quasars with $2.2<z<3$ are similar to those
of A and F stars, making selection of these objects difficult
\citep{fan99a,ric02,ric06}. X-ray and infrared surveys provide other ways to 
determine the QLF of both type I and type II AGNs 
\citep[e.g.][]{ued03,bar05,has05,bro06}. These studies have shown that a 
substantial fraction of AGNs are optically obscured at low luminosities. 
\citet{bar05} found a downturn at the faint end of the hard X-ray luminosity 
function for type I AGN, but current optically-selected quasar samples are 
not sufficiently faint to probe this downturn, if it exists.

To probe these issues, we need a large, homogeneous, faint quasar sample. 
This sample should be deep enough to study quasar behavior at the faint 
end, and large enough to provide good statistics. The sample should also 
span a large redshift range, straddling the peak of quasar activity.
All these require a wide-field spectroscopic survey of faint quasars selected
from deep multi-color imaging data. Such data are provided by the SDSS 
Southern Survey, a deep survey based on repeated imaging of the Fall 
Celestial Equatorial Stripe in the Southern Galactic Cap (SGC). The SGC 
imaging data, when co-added, reach more than two
magnitudes deeper than does the SDSS main survey, allowing efficient
selection of much fainter quasar candidates. The goal of the SDSS faint 
quasar survey (SFQS) is to obtain more than 1000 faint quasars from 10 
deg$^2$ of the SDSS deep stripe. This paper presents the first results of 
the SFQS, in an area of $\sim3.9$ deg$^2$. The spectroscopic observations 
were performed on MMT/Hectospec \citep{fab98}. The preliminary faint quasar 
sample reaches $g=22.5$, and fills in the crucial gap between large-area, 
shallow surveys such as the SDSS and 2QZ,
and deep, pencil beam surveys such as the $Chandra$ Deep Field (CDF) survey 
\citep[e.g.][]{bar05} and COMBO-17 survey \citep[e.g.][]{wol03}.

In $\S$ 2 of this paper, we introduce the SDSS deep imaging survey and 
the color selections of faint quasars from the deep data. In $\S$ 3,
we describe follow-up spectroscopic observations on MMT/Hectospec, and 
present the preliminary sample of the SFQS. We derive the QLF for the SFQS 
sample in $\S$ 4, and compare our results to other surveys in $\S$ 5.
Throughout the paper we use a $\Lambda$-dominated flat cosmology with 
H$_0$ = 70 km s$^{-1}$ Mpc$^{-1}$, $\Omega_{m}$ = 0.3, and 
$\Omega_{\Lambda}$ = 0.7 \citep[e.g.][]{spe03}.

\section{Quasar selection in the SDSS Southern Survey}

\subsection{The SDSS Southern Survey}

The Sloan Digital Sky Survey (SDSS) is an imaging and spectroscopic survey 
of the sky \citep{yor00} using a dedicated wide-field 2.5 m telescope 
\citep{gun06} at Apache Point Observatory, New Mexico. Imaging is carried
out in drift-scan mode using a 142 mega-pixel camera \citep{gun98} which
gathers data in five broad bands, $ugriz$, spanning the range from 3000 
to 10,000 \AA{ } \citep{fuk96}, on moonless photometric \citep{hog01} 
nights of good seeing.
The images are processed using specialized software \citep{lup01,sto02}, 
and are photometrically \citep{tuc06} and astrometrically \citep{pie03} 
calibrated using observations of a set of primary standard stars \citep{smi02} 
on a neighboring 20-inch telescope. The photometric calibration is accurate
to roughly 2\% rms in the $g$, $r$, and $i$ bands, and 3\% in $u$ and $z$,
as determined by the constancy of stellar population colors \citep{ive04,bla05},
while the astrometric calibration precision is better than 0.1 arcsec rms per
coordinate \citep{pie03}. All magnitudes are roughly on an AB system 
\citep{aba04}, and use the asinh scale described by \citet{lup99}.
From the resulting catalogs of objects,
complete catalogs of galaxies \citep{eis01,str02} and quasar candidates
\citep{ric02} are selected for spectroscopic follow-up \citep{bla03}.
Spectroscopy is performed using a pair of double spectrographs with coverage
from 3800 to 9200 \AA{ }, and a resolution $\lambda/\Delta \lambda$ of 
roughly 2000. The SDSS main quasar survey targets quasars 
with $i<19.1$ at $z\la 3$ and $i<20.2$ at $z\ga 3$ \citep{ric02}.
Its spectroscopic survey
is based on its imaging data with an exposure time of 54 seconds, so it 
is a shallow survey, and can only sample the most luminous end of the QLF.

In addition to the main imaging survey, the SDSS also conducts a deep imaging
survey, the SDSS Southern Survey, by repeatedly imaging the Fall Celestial
Equatorial Stripe in the Southern Galactic Cap. When completed, the
270 deg$^2$ area will be
imaged up to 30 times. The multi-epoch images, when co-added, allow the 
selection of much fainter quasar candidates than the SDSS main survey.

\subsubsection{Co-added catalog}

Quasar candidates are selected from the co-added catalog of the SDSS deep
stripe, i.e., each run goes through the photometric pipeline separately,
and the resulting catalogs are co-added. A better way to use multi-epoch 
images for quasar selection is to use co-added images, 
instead of multi-epoch catalogs. At the time when the spectroscopic
observations were carried out, co-added images were not available, so
in this paper quasar candidates were selected from the co-added catalog.

To construct the co-added catalog, we matched the multi-epoch data 
against themselves using a $0\farcs5$ tolerance. Given
$N_{epoch}$ 
epochs for a given object, a proper co-addition requires that we transform 
from $asinh$ magnitudes \citep{lup99} into flux. For a given SDSS band $j$, 
the conversion of magnitude $m_j$ into flux $f_j$ in Jy is given by
\begin{equation}
  f_j = 2 F_0 L_j \sinh \left [ -m_j/P - \ln L_j \right ],
  \label{eq:mag2flux}
\end{equation}
where $F_0=3630.78$ Jy, $P=1.08574$ and
$L=[1.4,0.9,1.2,1.8,7.4 ] \times 10^{-10}$ for the $u$, $g$, $r$, $i$, and $z$ 
bands, respectively \citep{sto02}. We then take the mean of the flux
$\bar{f}_j$ from the $N_{epoch}$ epochs and use the inverse of
Equation~\ref{eq:mag2flux} to recover the co-added magnitude $\bar{m}_j$.
For the error on the co-added magnitude, we calculate the standard deviation of
the fluxes from the epoch data $\Delta f_j$ and convert it to a magnitude
error $\Delta m_j$ using
\begin{equation}
  \Delta m_j = \frac{P \Delta f_j}{2 F_0}
  \left [ \sinh^2 \left(-\bar{m}_j/P - \ln L_j \right) +
    L_j^2 \right ]^{-\frac{1}{2}}.
\end{equation}

Figure 1 gives $u-g$, $g-r$ color-color diagrams for point sources 
($20.5<g<21.0$) in the SDSS main survey (single-epoch data) and the deep 
survey (multi-epoch data) with $N_{epoch}\sim13$. Each panel in Figure 1 
includes 10,000 objects.
Compared to the main survey, the stellar locus in the deep survey is much more
concentrated due to the smaller photometric errors, and $UVX$ quasar candidates
(confined by solid lines; see $\S$ 2.2) are well separated from the stellar 
locus. This enables us to
improve the quasar candidate selection and select much fainter quasars in 
color-color diagrams.

Figure 2 compares the magnitude limit and area of the SFQS with the 
LBQS \citep{fol87}, 2QZ \citep{boy00,cro04}, SDSS \citep{ric02}, 
2SLAQ \citep{ric05}, COMBO-17 \citep{wol03} and the CDF \citep{bar05} 
surveys of quasars and AGNs. The SFQS goes $\ga 2$ magnitude
deeper than 2QZ and SDSS, reaching the traditional quasar/AGN boundary at 
$z\sim2.5$, the peak of luminous quasar density evolution. It fills in the
crucial gap between large-area, shallow surveys such as the SDSS,
and deep, pencil beam surveys such as the CDF survey.

\subsection{Quasar candidate selection}

Quasar candidates are selected as outliers from the stellar locus in
color-color diagrams \citep[e.g.][]{new97}. The SDSS selects
quasar candidates based on their morphology and non-stellar
colors in $ugriz$ broad bands. The loci of simulated quasars
and Galactic stars in the SDSS $ugriz$ space are given in \citet{fan99a},
and the quasar color-selection in the SDSS is addressed in detail in
\citet{ric02}. In the SDSS main quasar survey, stellar outliers are defined
as those more than $4\sigma$ from the stellar locus in the $u-g$, $g-r$, 
$r-i$ and $g-r$, $r-i$, $i-z$ 3-D color spaces \citep{ric02}. We modify 
the SDSS selection criteria and select quasar candidates in 2-D
color-color diagrams in the SFQS survey. First, we generate a database
of simulated quasars in different redshift ranges \citep{fan99a,ric06}.
We make sure that our selection criteria can recover a
substantial fraction of the simulated quasars in each redshift ranges, 
including the mid-$z$ range. The photometric errors increase as quasar 
candidates go fainter, so we use slightly different selection criteria
in different magnitude ranges, and find a compromise between completeness and 
efficiency. There is a small amount of bright candidates that were already
observed spectroscopically in the SDSS main survey, and we do not observe
them in the SFQS. The final consideration is the fiber density of 
MMT/Hectospec (see $\S$ 3.1). The candidate density (excluding those that
already have spectra from the SDSS) in the sky is set to be $\sim$15\%
larger than the fiber density used for the quasar survey, so that
every fiber will be used in the case that candidates are closer than
the separation (20$\arcsec$) of adjacent fibers.

The spectroscopic observations were carried out on the 6.5m MMT
with Hectospec in June 2004 (hereafter Run I) and November 2004 (hereafter 
Run II). Run I was a pilot run, and used to test the integration
time, target selection criteria, and the data reduction software. We then
adjusted the integration time, and improved the quasar selection in Run II 
based on the observations in Run I, so the selection criteria in the two
runs were different. In the following subsections we mainly discuss the color
selection in Run II, and briefly in Run I where the selection was different. 
The Run II candidates were selected from the co-added catalog with average 
epoch number $N_{epoch}\sim13$. We restricted ourselves to objects with
$g>16.0$ selected from regions with $N_{epoch}>7$. We only 
selected point candidates in the two runs.

\subsubsection{$UVX$ and mid-$z$ candidates}

The ultraviolet excess ($UVX$) and mid-$z$ quasar candidates were selected
using $u-g$, $g-r$ diagrams. In Run II, the $g$ magnitude limit was 22.5.
For different magnitude ranges, we used slightly different color cuts.
As candidates go fainter, $\sigma$ increases, and the loci of stars in 
color-color diagrams become less concentrated. So our selection regions
at fainter ranges are a
little further from the stellar loci to reduce the contamination from stars.
The selection is summarized in Figure 3. The regions confined by solid lines
in Figure 3 are our selection regions. The left-hand box in each panel defines
$UVX$ candidates, and the right-hand box defines mid-$z$ candidates. 
In the top panels of Figure 3 where objects are bright, quasar candidates are
well separated from Galactic stars. But in the lower panels where objects 
are much fainter, quasars and stars blend together heavily.
In this case, we find a compromise between completeness and efficiency so that
the selection can recover a large fraction of simulated quasars, and the
candidate density exceeds the Hectospec fiber density used for our
quasar survey by $\sim$ 15\%.

In each panel of Figure 3, the right-hand selection box is used to 
recover a fraction of mid-$z$ quasars ($2<z<3$). The colors of mid-$z$
quasars are similar to those of stars (mainly late A and early F 
stars), so in $u-g$, $g-r$ diagrams, the locus of mid-$z$ quasars
partially overlaps the stellar locus \citep{fan99a,ric02}. 
The SDSS main survey selects mid-$z$ candidates using the selection 
similar to that shown in Figure 3, however, 
it only targets bright objects, and its mid-$z$ selection box is 
overwhelmed by contaminant stars. To limit the reduction in efficiency,
the main survey targets on 10\% of the objects in this 
selection box \citep{ric02}. In the SFQS, we target all mid-$z$ candidates
with acceptable efficiency, because contaminant A and F stars in the 
mid-$z$ selection box become less abundant at fainter magnitudes.
Figure 3 shows that the number of A and F stars drops rapidly at 
$g\sim20.5$, due to the fact that we have reached the most distant early F
dwarfs in the Galactic halo at this magnitude. 

In addition to the candidates selected by Figure 3, we obtain another 
candidate sample down to $g=22.0$ independently from the co-added catalog
using the kernel density 
estimator \citep[KDE;][]{sil86} technique described by \citet{ric04} who 
applied this method to the SDSS-DR1 imaging area. The KDE 
method \citep{gra04} is a 
sophisticated extension of the traditional color selection technique for 
identifying quasars \citep[e.g.][]{ric04}. For our case, 
we have applied the algorithm to data that is considerably fainter than was 
used by \citet{ric04} ($g=22.0$ as compared to $g=21.0$). While the SDSS 
imaging is complete to this depth, the errors are larger than is ideal for the
application of this method.

Our final candidate sample is the combination of the two independent samples.
In fact, the two samples contain roughly the same candidates at $g<22.0$.
In Run II, there are only 15 quasars included by the KDE sample but not 
included by the selection in Figure 3.

The candidate selection in Run I was slightly different: (1) In Run I, the
average epoch number of the co-added catalog was 7.4, so the photometric 
errors were larger than those in Run II; (2) The selection in
Run I was based on $r$ magnitude, and the selection of $UVX$ and mid-$z$ 
candidates was down to $r=22.5$ (but the selection efficiency is only 10\% at
$22.0<r<22.5$, see $\S$ 3), and the bright limit was $r>15.5$;
(3) In Run I we did not use the mid-$z$
selection box in the first panel of Figure 3
($r<20.0$), which means that we missed bright mid-$z$ quasars in Run I;
(4) We did not use the KDE method to obtain an independent candidate sample.
Therefore the selection efficiencies and incompleteness are different for the
two runs, and we will correct their incompleteness separately.

\subsubsection{High-$z$ candidates}

The color selection of high-$z$ quasar candidates in the SDSS color space is 
well studied in a series of papers by \citet{fan99b,fan00,fan01a}. Similar to
\citet{ric02}, we define three regions for high-$z$ quasars in $ugr$, $gri$,
$riz$ color space. They are used to recover quasars at $z>$ 3.0, 3.6, and 
4.5 respectively. When $z>3.7$, the Ly$\alpha$ line enters the $r$ band, 
so the selection of high-$z$ candidates is based on the $i$ magnitude and 
down to $i=22.5$. Again, we use slightly different selection criteria for 
different $i$ magnitude ranges, due to increasing photometric errors with 
decreasing brightness.

\section{Observation and data reduction}

\subsection{Spectroscopic observation and data reduction}

The spectroscopic survey of faint quasars was carried out with the 6.5m MMT 
with Hectospec \citep{fab98}.
Hectospec is a multiobject optical spectrograph with 300 fibers, and a
$1\degr$ field of view. With a 270 line mm$^{-1}$ grating, Hectospec covers a
wavelength range of 3700 to 9200 \AA { } at a moderate resolution of $\sim6$
\AA. This is sufficient to measure the redshifts of quasars at any redshift 
lower than 6, and provide robust line-width measurement.

Simultaneously with the faint quasar survey, we also conducted 
a spectroscopic survey of luminous early-type galaxies in the same SDSS 
fields. The early-type galaxy survey is described in detail in 
\citet{coo06}.
We divided Hectospec fibers equally between quasar and galaxy targets. For
each configuration, 30 sky fibers and 5 F subdwarf standard stars were used
for calibration, and approximately 130 fibers were used to target quasar
candidates. In 2004, five Hectospec fields in Runs I and II were observed.
The central position and exposure time for each field are given in Table 1.

The Hectospec data were reduced using HSRED, an IDL package developed for
the reduction of data from the Hectospec \citep{fab98} and Hectochelle 
\citep{sze98} spectrographs on the MMT, and based heavily on the reduction 
routines developed for processing of SDSS spectra \citep{sch06}.
                                                                                
Initially, the two dimensional images are corrected for cosmic ray
contamination using the IDL version of L.A. Cosmic \citep{dok01} developed
by J. Bloom. The 300 fiber trace locations
are determined using dome flat spectra obtained during the same night as each
observation; the CCD fringing and high frequency flat fielding variations are
also removed using these dome flats. On nights when twilight images are
obtained, these spectra are used to define a low-order correction to the flat
field vector for each fiber.
                                                                                
Each configuration generally includes approximately 30 fibers located on blank
regions of the sky.
Using the bright sky lines in each spectrum, we adjust the initial
wavelength solution, determined from HeNeAr comparison spectra, to compensate
for any variations throughout the night.  These sky lines are used further to
determine any small amplitude multiplicative scale offset for each fiber,
occurring due to small variations in the relative transmission differences
between fibers not fully corrected using the flat-field spectra,  before the
median sky spectrum is subtracted.
                                                                                
The data are fluxed using SDSS calibration stars observed on the same
configuration as the objects of interest.  These F stars are cross-correlated
 against a grid of \citet{kur93} model atmospheres to determine the best fit
stellar spectrum.  SDSS photometry of the standard stars is then used to
determine the absolution normalization of the standard star spectrum. The
ratio of this master spectrum and the observed count rate is used to construct
the fluxing vector for each exposure.  After each exposure is extracted,
corrected for heliocentric motion and flux calibrated, the spectra are
de-reddened according galactic dust maps \citep{sch98} with the
\citet{don94} extinction curve. Finally, multiple exposures of a single
field are combined to obtain the final spectrum. After 180 minute exposure
on a $g=22.0$ object, the typical signal-to-noise ratio 
at $\sim 5000$ \AA { } reaches $\sim7$ per pixel.

Redshifts are determined using programs available in the IDLSPEC2D package of
IDL routines developed for the SDSS. For each observed spectrum, the best fit
spectral template and redshift are obtained from a number of quasar, galaxy, 
and star spectra using $\chi^{2}$ fits. The Hectospec has sufficient 
wavelength coverage for reliable redshift measurement. The success rate is 
better than 90\% for $g<22.0$. After the automatic identification, each 
redshift is examined by eye to guard against failed redshifts or
misclassifications. Quasar identification is not easy for
faint candidates with $g>22.0$, especially when they also have
weak emission lines. We correct the fraction of unidentifiable objects
statistically.

As we mentioned in $\S$ 2.2, the quasar candidate density in the sky is
set to be $\sim$ 15\% larger than the fiber density used for quasar survey,
so we did not observe all candidates in the fields. We will correct the
incompleteness arising from this fact in $\S$ 4.1.

\subsection{Faint quasar sample}

The preliminary sample of the SFQS consists of 414 quasars from 5 Hectospec 
fields ($\sim3.9$ deg$^2$). The sample has a redshift range from 0.32 to 4.96,
with 119 objects at $z>2.0$ and 23 objects at $z>3.0$. The median value of the 
redshifts is 1.72, greater than the median redshift 1.47 in the SDSS DR3
\citep{sch05}. Most of the non-quasar candidates are 
star-forming galaxies, A stars, and white dwarfs (WDs). For example, in Run II,
non-quasar candidates consist of $\sim 40\%$ star-forming galaxies, 
$\sim 30\%$ A or early type stars, $\sim 20\%$ WDs (including M star-WD
pairs), and $\sim 10\%$ M or late type stars. Most of the A stars are from the 
mid-$z$ quasar selection. The loci of these contaminant objects in color-color
space, and why they are selected as quasar candidates are well addressed
in \citet{fan99a} and \citet{ric02}.

Figure 4 illustrates the numbers of candidates observed and the selection 
efficiencies in the two runs. In Figure 4, the black areas are unidentifiable
objects, the gray areas are identified as non-quasars, and the blank regions
are identified as quasars. The fractions of quasars are also given within or 
above the bars. The total selection efficiency in Run II is $\sim 66\%$, and 
the efficiency at $g<21.5$ is as high as 79\%. Much of the low efficiency
is produced by the mid-$z$ candidates, where the total efficiency and the 
efficiency at $g<21.5$ are 43\% and 35\%, respectively. 
In Run I, the selection efficiency at $r>22.0$ is only 10\%. 
The average epoch number $N_{epoch}$ of the data in Run I is 7.4,
smaller than 13.0 in Run II, so the photometric errors are relatively
larger, and the quasar selection is thus less efficient.
Due to the low efficiency at $r>22.0$, the quasar selection in Run I is only
complete to $r=22.0$. In Run II, the quasar selection probes to $g=22.5$.
Note that for a quasar with a power-law continuum slope
of $-0.5$ (see $\S$ 3.3), its $g$ magnitude is fainter than $r$ magnitude
by 0.15. We also improved the selection criteria in Run II based on Run I (see
$\S2.2$), so the selection efficiency in Run II was increased. 

Figure 5 gives the $i$ magnitude and redshift distributions of the SFQS
sample. The dashed profiles are from the SDSS DR3 \citep{sch05}, and have 
been scaled to compare with our survey. The SFQS sample is about $2\sim3$ 
magnitudes fainter than the SDSS main survey as we see in the left panel. 
In the right panel, our survey peaks at a similar redshift $z\sim1.6$ to
the SDSS main survey, but contains a larger 
fraction of $2<z<3$ quasars due to our more complete selection of quasar
candidates in this redshift range. There is a small dip at $z\approx 2.7$.
It may be caused by the fact that we missed bright mid-$z$ quasars in Run I. 

Figure 6 shows six sample spectra obtained by our survey: (a) A typical 
bright quasar with $g=20.11$; (b) A typical faint quasar with $g=22.36$;
(c) A typical low-$z$ quasar at $z=0.56$; (d) The most distant quasar observed
in the two runs at $z\sim5$; (e) A high-$z$ quasar with a broad CIII] emission
line and a series of narrow emission lines such as Ly$\alpha$, and CIV.
\citep{zak03}; (f) A broad absorption line quasar at $z=2.02$.

Table 2 presents the quasar catalog of the SFQS. Column 1 gives the name
of each quasar, and column 2 is the redshift. Column 3 and 4 list the
apparent magnitude $g$ and the rest-frame absolute magnitude $M_g$.
The slope $\alpha$ of the power-law
continuum for each quasar is given in column 5. Measurements of $M_{g}$ and
$\alpha$ are discussed in the next subsection. The full Table 2 will appear
in the electronic edition.

\subsection{Determination of continuum properties}

We determine the continuum properties for each quasar using the observed
spectrum and the SDSS photometry. Hectospec gives spectra from $\sim$ 3700
to $\sim$ 9200 \AA; however, at the faint end of the sample, the observed 
spectra have low signal-to-noise ratios and could be strongly effected by
errors in 
flux calibration or sky subtraction. The accurate broadband photometry of the 
SDSS provides us an alternative way to determine the continuum 
properties from both the observed spectrum and the SDSS photometry 
\citep{fan01b}. For a given quasar, we obtain the intrinsic 
spectrum by fitting a model spectrum to the broadband photometry. The model
spectrum is a power-law continuum plus emission lines. For the power-law 
continuum $f_{\nu}=A\times{\nu}^{\alpha}$, we do not assume a uniform 
slope $\alpha=-0.5$, instead, we determine the slope and the normalization 
$A$ for each quasar. To obtain model emission lines, we measure the strength
of the observed emission line with the highest signal-to-noise ratio. The
strengths of other emission lines are determined using the line strength 
ratios from the composite spectrum given by \citet{van01}.

The SDSS magnitudes $m^{model}$ for the model spectrum are directly 
calculated from the model spectrum itself. Then $\alpha$ and $A$ are 
determined by minimizing the differences between the model magnitudes 
$m^{model}$ and the SDSS photometry $m^{obs}$,
\begin{equation}
  {\chi}^2 = \sum(\frac{m^{model}_{i} - m^{obs}_{i}}{\sigma^{obs}_{i}})^{2},
  \label{eq:mindiff}
\end{equation}
where $\sigma^{obs}_{i}$ is the SDSS photometry error. We fix the value of 
$\alpha$ in the range of $-1.1<\alpha<0.1$. In Equation~\ref{eq:mindiff}, 
we only use the bands that are not dominated by Ly$\alpha$ absorption systems.
With the information of intrinsic spectra and redshifts, we calculate the 
absolute magnitudes. The slope $\alpha$ and absolute magnitude $M_{g}$ are 
given in Table 2.

\section{Optical luminosity function of faint quasars}

In this section, we correct for the photometric, 
coverage, and spectroscopic incompleteness and the morphology bias.
We then use the traditional $1/V_{a}$ method \citep{avn80} to derive a binned
estimate of the luminosity function for the SFQS sample and model
the luminosity function using maximum likelihood analysis.

\subsection{Completeness corrections}

The photometric incompleteness arises from the color
selection of quasar candidates. It is described by the selection function,
the probability that a quasar with a given magnitude, redshift, and intrinsic
spectral energy distribution (SED) meets the color selection criteria
\citep[e.g.][]{fan01b}. By assuming that the intrinsic SEDs have certain 
distributions, we can calculate the average selection probability as a 
function of magnitude and redshift. To do this, we first calculate the 
synthetic distribution of quasar colors for a given ($M_g,z$), following the 
procedures in \citet{fan99a}, \citet{fan01b} and \citet{ric06}.
Then we calculate the SDSS 
magnitudes from the model spectra and incorporate photometric errors into 
each band. For an object with given ($M_g,z$), we generate a database of model
quasars with the same ($M_g,z$). The detection probability for this quasar is 
then the fraction of model quasars that meet the selection criteria.

Figure 7 gives the selection probabilities as the function of $M_g$ and $z$ 
in the two runs.
The contours are selection probabilities from 0.2 to 0.8 with an interval of 
0.2, and heavy lines (probability $=0.2$) illustrate the limiting magnitudes.
The solid circles are the locations of sample quasars. The two selection 
functions are different due to the different color selection criteria used.
The striking difference is the detection probabilities in the mid-$z$ range, 
where quasars are difficult to select by their SDSS colors. In 
Run I, the probabilities in the mid-$z$ range are very low for luminous
quasars. But in Run II, the selection in this range is greatly 
improved. This makes an almost homogeneous selection from $z=0$ to $z>5$.
Due to this improved selection, we are able to correct for the incompleteness
down to $M_g=-22.0$, $-23.0$, $-24.0$ at $z<2$, $z<3$, and $z<4$, respectively.

The second incompleteness is the coverage incompleteness, which comes from
the fact that we did not observe all candidates in the fields due to the 
limited fiber density of Hectospec. To correct this effect, we assume
that the selection efficiency of unobserved candidates is the same as that
of observed ones.

The third incompleteness, spectroscopic incompleteness, comes from the fact
that we cannot identify
some faint candidates due to their weak flux observed on Hectospec. 
To correct this incompleteness, we assume that unidentifiable candidates 
have the same selection efficiency of identified ones with the same 
magnitudes. From the two runs we know that, with the capacity of Hectospec and
the integration time of $\sim180$ minutes, the emission lines of a typical 
broad-line quasar with $g\sim22.5$ should be visible in the Hectospec spectra,
so unidentifiable candidates are either weak-line quasars, or not quasars.
Therefore this correction may give an upper limit.

Another incompleteness arises from the morphology bias. The candidates we 
observed are point sources, but faint point sources could be mis-classified
as extended sources by the SDSS photometric pipeline \citep[e.g.][]{scr02}.
The best way to correct this incompleteness is to observe a sample of extended
sources that satisfy our selection criteria, and find the fraction of 
quasars among them. The definition of star-galaxy classification in the SDSS 
photometric pipeline gives us an alternative way. 
In the SDSS photometric pipeline, an object is defined as a $galaxy$
(extended object) if $psfMag-cmodelMag>0.145$,
where $psfMag$ is the PSF magnitude, and $cmodelMag$ is the composite model
magnitude determined from the best-fitting linear
combination of the best-fitting de Vaucouleurs and exponential model for an
object's light profile \citep{aba04}. Similar to \citet{scr02}, we define the
difference between $psfMag$ and $cmodelMag$ as $concentration$.
To correct the morphology bias, we plot the $concentration$ distribution 
vs. object counts as shown in Figure 8, where the dash-dotted lines separate 
$stars$ and $galaxies$ by definition. At $g>22.0$, the star locus and the 
galaxy locus begin 
to mix heavily, and the objects near the separation lines may be 
mis-classified. To estimate the real numbers of point and extended sources, 
we use double Gaussians to fit each component of the $star$ and $galaxy$ 
loci as shown in the figure. Then the fraction of point sources 
misclassified as extended ones are obtained from the best fits.
They are 10\% for $22.0<g<22.2$, 24\% for $22.2<g<22.4$, and 
31\% for $22.4<g<22.5$.

In Run I and Run II we only select point sources. Although we have corrected
the morphology bias, our sample could still be biased by not including objects
in which the host galaxies are apparent. So the low-$z$ QLF at the faint end
may be affected by quasar host galaxies. We plan to observe a sample of
extended sources, and correct the morphology bias and the effect of host
galaxies in the next observing run.

\subsection{1/$V_{a}$ estimate and maximum likelihood analysis}

We derive a binned estimate of the luminosity function for the SFQS sample 
using the traditional $1/V_{a}$ method \citep{avn80}. The available volume 
for a quasar with absolute magnitude $M$ and redshift $z$ in a magnitude bin 
$\Delta M$ and a redshift bin $\Delta z$ is 
\begin{equation}
V_{a} = \int_{\Delta M}\int_{\Delta z}p(M,z) \frac{dV}{dz} dzdM,
\end{equation}
where $p(M,z)$ is a function of magnitude and redshift and used to 
correct sample incompleteness. Then the luminosity function and 
the statistical uncertainty can be written as
\begin{equation}
\Phi(M,z) = \sum \frac{1} {V_{a}^{i}}, \ \
\sigma(\Phi) = \left[\sum (\frac{1} {V_{a}^{i}})^2 \right]^{1/2},
\end{equation}
where the sum is over all quasars in the bin. This is essentially the same as
the revised $1/V_{a}$ method of \citet{pag00}, since $p(M,z)$ has already 
corrected the incompleteness at the flux limits.

The QLF derived from the $1/V_{a}$ estimate is shown in Figure 9, which gives 
the QLF from $z=0.5$ to 3.6, with a redshift interval of $\Delta z\approx 0.5$.
In each redshift bin, the magnitude bins are chosen to have exactly the 
same numbers of quasars except the brightest one. Solid symbols represent 
the QLF corrected for all four incompletenesses in $\S 4.1$, and open symbols 
represent the QLF corrected for all incompletenesses
except the spectroscopic incompleteness. Our
sample contains two subsamples from Run I and Run II. The two subsamples are 
weighted by their available volumes in each magnitude-redshift bin when 
combined. In Run I, the selection efficiency at $r>22.0$ is only 10\%, so we
do not include the quasars of $r>22.0$ in the 1/$V_a$ estimate.
We also exclude the Run I quasars in the brightest bins at $2.0<z<3.0$,
due to the low selection completeness in this range.

Our sample contains a small fraction of luminous quasars.
By comparing the SFQS QLF with the results of the 2QZ, 2SLAQ, and SDSS
(see Figure 11 and Figure 12 in $\S 5.3$), one can see that the QLF
has steep slopes at the bright end (from the 2QZ, 2SLAQ, or SDSS) and much 
flatter slopes at the faint end (from the SFQS), clearly showing a break
in the QLF, thus we will use the double power-law form to model the QLF.
At $z>2.0$, quasars show strong density evolution. A double power-law QLF 
with density evolution requires at least six parameters, however, this 
sample is not large enough to determine so many parameters simultaneously. 
Therefore we break the sample to two subsamples, $0.5<z<2.0$ and $2.0<z<3.6$,
and use fewer free parameters to model them separately, based on
reasonable assumptions. We will model the low-$z$ and high-$z$ quasars 
simultaneously when we obtain more than 1000 quasars in the future.

\subsubsection{QLF at 0.5 $<z<$ 2.0}

As the first step, we try a single power-law form $\Phi(L)\propto L^{\beta}$
to model the observed QLF. The slope $\beta$ determined from the best fit is 
$-1.40$, much flatter than the bright-end slopes ($-3.0 \sim -3.5$)
of the QLFs from the 2QZ \citep{boy00,cro04}, SDSS \citep{ric06}, and 2SLAQ 
\citep{ric05}. This confirms the existence of the break in the slope.
We also use the single power-law form to model the three individual redshift 
bins. The best fit slopes at $0.5<z<1.0$, $1.0<z<1.5$, and $1.5<z<2.0$ are 
$-1.45_{-0.3}^{+0.2}$, $-1.40_{-0.2}^{+0.1}$, and $-1.35_{-0.2}^{+0.1}$,
respectively. They are consistent within $1\sigma$ level, so 
there is no strong evolution in the slope at the faint end.

To characterize the QLF at $0.5<z<2.0$, we use the double power-law form with
PLE, which expressed in magnitudes is,
\begin{equation}
\Phi_{L}(M_{g},z)=\frac{\Phi(M_{g}^{\ast})}{10^{0.4(\alpha+1)(M_{g}-
  M_{g}^{\ast}(z))}+10^{0.4(\beta+1)(M_{g}-M_{g}^{\ast}(z))}}.
\end{equation}
In the case of PLE models, the 
evolution of the characteristic magnitude $M_{g}^{\ast}(z)$ can be modeled as
different forms, such as a second-order polynomial
evolution $M_{g}^{\ast}(z)=M_{g}^{\ast}(0)-2.5(k_{1}z+k_{2}z^{2})$,
an exponential form $M_{g}^{\ast}(z)=M_{g}^{\ast}(0)-1.08k\tau$,
where $\tau$ is the look-back time, or, a form of 
$M_{g}^{\ast}(z)=M_{g}^{\ast}(0)-2.5k(1+z)$ \citep[e.g.][]{boy00,cro04}.
\citet{cro04} has shown that low-$z$ QLF can be well fit by both second-order 
polynomial or exponential evolution in the $\Lambda$ cosmology. In this paper
the quasar sample is still small, so we do not fit all models and
justify their validity; instead, we take the exponential form, which uses fewer
parameters.

We use maximum likelihood analysis to find the best fits. The likelihood
function \citep[e.g.][]{mar83,fan01b} can be written as
\begin{equation}
S = -2\sum ln[\Phi(M_{i},z_{i})p(M_{i},z_{i})] + 2\int\int \Phi(M,z)p(M,z) 
    \frac{dV}{dz} dzdM,
\end{equation}
where the sum is over all quasars in the sample.
Our sample does not contain enough bright objects, which makes it difficult to
determine the slope at the bright end. We thus fix the 
bright-end slope to the value of $\alpha=-3.25$ given by \citet{cro04}.
The best fits are, $\alpha=-3.25$ (fixed), $\beta=-1.25_{-0.15}^{+0.10}$, 
$k=7.50_{-0.35}^{+0.40}$, $M_{g}^{\ast}(0)=-19.50_{-0.35}^{+0.40}$, and 
$\Phi(M_{g}^{\ast})=1.84\times 10^{-6}$ Mpc$^{-3}$ mag$^{-1}$. 
The $\chi^2$ of this fit is 15.0 for 15 degrees of freedom by comparing
the $1/V_{a}$ estimate and the model prediction.
The solid lines in Figure 9 are the best model fits. For 
comparison, the dashed lines are the best fit at $1.5<z<2.0$. 
One can see that the density of quasars increases from $z$ = 0.5 to 
2.0, then decreases at higher redshift.

As stated in $\S$ 4.1, the correction of the spectroscopic incompleteness
may only provide an upper limit. In Figure 9, solid symbols represent the QLF
corrected for all four incompleteness in
$\S 4.1$, and open symbols represent the QLF corrected for all incompleteness
except the spectroscopic incompleteness. One can see that the spectroscopic
incompleteness only affects the faintest bins. The best model fit shows that,
without the correction for the spectroscopic incompleteness, the slope at
$0.5<z<2.0$ flattens from $-1.25_{-0.15}^{+0.10}$ to $-1.10_{-0.15}^{+0.10}$.
The variation in the slope is within $1\sigma$ level.

\subsubsection{QLF at 2.0 $<z<$ 3.6}

At $z>2.0$, the QLF cannot be modeled by PLE, and density evolution is
needed. We add a density evolution term $\rho_{D}(z)$ into the double 
power-law form to model the QLF at $z>2.0$. The double power-law model with
density evolution expressed in magnitudes is,
\begin{equation}
\Phi_{H}(M_{g},z)=\frac{\Phi(M_{g}^{\ast})\rho_{D}(z)}{10^{0.4(\alpha+1)(M_{g}-
  M_{g}^{\ast}(z))}+10^{0.4(\beta+1)(M_{g}-M_{g}^{\ast}(z))}},
\end{equation}
where we take the exponential form for the evolution of characteristic 
magnitude $M_{g}^{\ast}(z)$ as we do for $z<2$, and take an exponential form
of $\rho_{D}(z)=10^{-B(z-2)}$ for the density evolution at a given magnitude 
\citep[e.g.][]{sch95,fan01b}. The single power-law model with density 
evolution shows that the QLF at $2.0<z<3.6$ has a slope of $-1.70$, flatter 
than the bright-end slopes ($-2.5 \sim -3.5$) from the COMBO-17 \citep{wol03}, 
SDSS \citep{ric06} and \citet{fan01b}. This indicates the existence of the 
break in the QLF at $2.0<z<3.6$.

We use the double power-law form with density evolution to model the observed
QLF at $2.0<z<3.6$.
As we do for $z<2.0$, we fix the bright-end slope $\alpha$ as
$-3.25$. The parameters $k$ and $M_{g}^{\ast}(0)$ are also fixed to the values
determined from $0.5<z<2.0$. There are three parameters $\beta$, $B$,
and $\Phi(M_{g}^{\ast})$ that we need to derive. $\Phi(M_{g}^{\ast})$
is not a free parameter, because Equations 6 and 8 must be consistent
at $z=2$. Figure 9 shows that, at $M_{g}\sim -23$, the density
at $1.5<z<2.0$ and the density at $2.0<z<2.5$ are roughly the same. So we
connect Equations 6 and 8 through
$\Phi_{L}(M_{g}=-23,z=2)=\Phi_{H}(M_{g}=-23,z=2)$.
Then $\Phi(M_{g}^{\ast})$ can be derived from $\beta$ and $B$ by this relation.
We use maximum likelihood analysis to determine the two free parameters
$\beta$ and $B$ as well as $\Phi(M_{g}^{\ast})$. 
The best fits are, $\beta=-1.55\pm 0.20$, $B=0.45 \pm 0.15$,
and $\Phi(M_{g}^{\ast})=1.02\times 10^{-6}$ Mpc$^{-3}$ mag$^{-1}$. 
The $\chi^2$ of this fit is 10.4 for 11 degrees of freedom.
The solid lines in Figure 9 are 
the best model fits.

Our sample contains only 5 quasars at $z>3.6$, which is lower than what 
we expected if the power-law slope of the QLF at $3.6<z<5.0$ is
$\sim -2.4\pm0.2$ \citep{ric06}. This result implies that the 
faint-end slope of the QLF at $3.6<z<5.0$ is also flatter than that at the
bright end.

\section{Discussion}

\subsection{Luminosity-dependent density evolution}

It is convenient to show the quasar evolution by plotting the space density
as a function of redshift. Our sample spans a large redshift range, covering
the mid-$z$ range with good completeness. Figure 10 gives the integrated
comoving density as the function of $z$ for three magnitude ranges, 
$M_g<-22.5$, $-23.5$, and $-24.5$, respectively. We use the redshift bins 
from $z=0.4$ to 3.6 with an interval of $\Delta z=0.4$. For the bins at 
$2.2<z<3.0$, we exclude quasars with $M_{g}<-24.5$ in Run I, due to the low 
selection completeness in this range. We also exclude incomplete bins in 
Figure 10. At low redshift, the space density steadily increases from 
$z\sim0.5$ to $\sim2$. Then it decreases toward high redshift. The quasar 
evolution peaks at $z\sim2$ in the range of $M_{g}<-22.5$.
Solid curves are integrated densities calculated from the best model fits in 
$\S$ 4.2, while dotted, dashed and dot-dashed curves represent the integrated
densities from the SDSS \citep{ric06}, 2QZ \citep{boy00}, and 
\citet[SSG]{sch95}, respectively.
X-ray surveys indicate that X-ray selected quasars and AGNs exhibit so-called
``cosmic downsizing'': luminous quasars peak at an earlier epoch in the cosmic
history than fainter AGNs \citep[e.g.][]{ued03,bar05,has05}. We cannot see the 
cosmic downsizing from the SFQS sample due to the small dynamical range in 
magnitude and large errors bars. However, the peak of $z\sim 2$ for the SFQS 
sample is later in cosmic time than the peak of $z\sim 2.5$ found from 
luminous quasar samples, such as the SDSS \citep{ric06}. 

\subsection{Comparison to other surveys}

In this section we compare the SFQS QLF with QLFs derived from the 2QZ 
\citep{boy00}, SDSS \citep{ric06}, 2SLAQ \citep{ric05}, COMBO-17 
\citep{wol03}, and CDF \citep{bar05}. The survey areas and magnitude limits
are sketched in Figure 2. Because different surveys use different cosmological
models, we convert their QLFs to the QLFs expressed in the cosmological model 
that we use. First, absolute magnitude $M'$ is converted to $M$ by
$M'-M=-5log_{10} \frac{d_{L}^{'}} {d_{L}}$,
where $d_{L}^{'}$ and $d_{L}$ are luminosity distances in different 
cosmologies. Then magnitudes in different wavebands are converted to $M_{g}$ 
in the same cosmology by
$M_{g}=M_{\lambda}+2.5\alpha log_{10} \frac {\lambda_{g}} {\lambda}$,
where $\alpha$ is the slope of the power-law continuum, and we assume 
$\alpha=-0.5$; $\lambda_{g}$ and $\lambda$ are the effective wavelengths of the 
two different wavebands. 
Finally spatial density $\rho^{'}$ in a $M$-$z$ bin is converted to $\rho$ by
$\rho=\rho^{'} \frac{V_{a}^{'}} {V_{a}}$,
where $V_{a}$ and $V_{a}^{'}$ are available comoving volumes in the different 
cosmologies.

Figure 11 gives the comparison with the 2QZ \citep{boy00} and 2SLAQ
\citep{ric05} at $0.5<z<2.0$. Compared to the 2QZ and 2SLAQ, the SFQS probes 
to higher redshifts and fainter magnitudes. Solid circles and open triangles in
Figure 11 are the SFQS QLF and 2SLAQ QLF, respectively. The dotted and dashed 
lines represent the 2QZ QLF, which is an average QLF calculated using
$<\Phi(M,z)>=\frac{\int_{\Delta z}\Phi(M,z)dV}{\int_{\Delta z}dV}$, where
$\Phi(M,z)$ is the best-fitting double power-law model with a second-order 
polynomial luminosity evolution from \citet{boy00}. The dashed-line parts
are roughly the range that the observed 2QZ QLF really covered \citep{boy00}.
The solid lines are the best model fits of the SFQS QLF. They give a good fit 
to all three QLFs. At the bright end of the QLF,
the SFQS, 2QZ, and 2SLAQ agree well. At the faint end, the 2QZ
predicts a higher density and a steeper slope than the SFQS.
By combining the deep SFQS with the 2QZ and 2SLAQ, one can see that
there is clearly a break in the QLF slope. 

Figure 12 shows the comparison with the SDSS \citep{ric06} and COMBO-17 
\citep{wol03} at $0.5<z<3.6$.
The SDSS is a shallow survey covering a large redshift range.
The COMBO-17 survey uses photometric redshifts, and collects quasars
in an area of $\la$1 deg$^2$. In Figure 12, solid circles and open triangles are
the SFQS QLF and SDSS QLF, respectively. The dashed lines represent the
COMBO-17 QLF, which is an average QLF calculated from the best model fitting
of \citet{wol03} using the same method that we did above. The solid lines
are the best model fits of the SFQS QLF, and they give a reasonable fit to both 
SDSS and SFQS QLFs. The SFQS is more than 2 magnitudes deeper than the SDSS, but
their QLFs are consistent at all redshifts. The combination of the two 
QLFs also shows the existence of a break in the slope. At $1.0<z<3.6$,
the COMBO-17 QLF agrees well with the SFQS QLF, although it has a flatter 
slope than the SDSS at the bright end.

Figure 13 shows the comparison with the CDF survey of quasars and AGNs 
\citep{bar05}. \citet{bar05} determine the luminosity functions for both type I
and type II AGNs selected from hard X-ray surveys, and find a downturn at the 
faint end of type I AGN luminosity function. We convert absolute magnitudes 
and X-ray luminosities to bolometric luminosities using the method given by
\citet{bar05}, and compare the SFQS QLF (solid circles) with the type I AGN hard 
X-ray luminosity function (open triangles and squares) of \citet{bar05} in 
Figure 13. At the bright end of the QLF, the two surveys are consistent at all 
redshifts. At the faint end, they agree well at $1.5<z<2.0$, 
however, at $0.5<z<1.5$, \citet{bar05} has significantly higher 
($\sim2\sigma$) densities. One can also see that the SFQS does not probe faint 
enough to reach the turndown seen by \citet{bar05}.

\section{Summary}

This paper presents the preliminary results of a deep spectroscopic survey of 
faint quasars selected from the SDSS Southern Survey, a deep imaging survey, 
created by repeatedly scanning a 270 deg$^2$ area. Quasar candidates are 
selected from the co-added catalog of the deep data. With an average epoch 
number $N_{epoch}\sim13$, the co-added catalog enables us to select much 
fainter quasars than the quasar spectroscopic sample in the SDSS main survey. 
We modify SDSS color selection to select quasar candidates, so that they cover
a large redshift range at $z<5$, including the range of $2<z<3$ with good 
completeness. Follow-up spectroscopic observations were carried out on 
MMT/Hectospec in two observing runs. With the capacity of Hectospec, 
the selection efficiency of faint quasars in Run II is $\sim$80\% at
$g<21.5$, and acceptable at $g>21.5$ (60\% for $21.5<g<22.0$ and 40\% for 
$22.0<g<22.5$). The preliminary sample of the SFQS contains 414 quasars and 
reaches $g=22.5$.

We use the $1/V_{a}$ method to derive a binned estimate of the QLF. By 
combining the SFQS QLF with the QLFs of the 2QZ, 2SLAQ, and SDSS, we conclude
that there is a break in the QLF. We use the double power-law form with PLE to 
model the observed QLF at $0.5<z<2.0$, and the double power-law form with an
exponential density evolution to model the QLF at $2.0<z<3.6$. The QLF slopes 
at the faint end ($-1.25$ at $0.5<z<2.0$ and $-1.55$ at $2.0<z<3.6$) are much 
flatter than the slopes at the bright end, indicating the existence of the 
break at all redshifts probed. The luminosity-dependent density evolution 
model shows that the quasar evolution at $M_{g}<-23.5$ peaks at $z\sim 2$, 
which is later in the cosmic time than the peak of $z\sim 2.5$ found from 
luminous quasar samples.

Our survey is compared to the 2QZ \citep{boy00}, SDSS \citep{ric06}, 2SLAQ 
\citep{ric05}, COMBO-17 \citep{wol03}, and CDF \citep{bar05}. The SFQS QLF is
consistent with the results of the 2QZ, SDSS, 2SLAQ and COMBO-17. The SFQS 
QLF at the faint end has a significantly lower density at $0.5<z<1.5$ than 
does the CDF. The preliminary sample of the SFQS is still small, and 
statistical errors are large. We plan to obtain more faint quasars
from future observations and establish a complete quasar sample with more 
than 1000 quasars over an area of 10 deg$^2$.

\acknowledgments

We would like to thank A. J. Barger for providing useful data, and A. G. 
Gray, B. Nichol, and R. Brunner for helping in the kernel density estimator 
technique. We thank the Hectospec instrument team and MMT staff for
their expert help in preparing and carrying out the Hectospec observing
runs. We acknowledge 
support from NSF grant AST-0307384, a Sloan Research Fellowship
and a Packard Fellowship for Science and Engineering (L.J., X.F.), and NSF 
grant AST-0307409 (M.A.S.).

Funding for the SDSS and SDSS-II has been provided by the Alfred P. Sloan 
Foundation, the Participating Institutions, the National Science Foundation, 
the U.S. Department of Energy, the National Aeronautics and Space 
Administration, the Japanese Monbukagakusho, the Max Planck Society, and the 
Higher Education Funding Council for England. 
The SDSS Web Site is http://www.sdss.org/.
The SDSS is managed by the Astrophysical Research Consortium for the 
Participating Institutions. The Participating Institutions are the American 
Museum of Natural History, Astrophysical Institute Potsdam, University of 
Basel, Cambridge University, Case Western Reserve University, University of 
Chicago, Drexel University, Fermilab, the Institute for Advanced Study, the 
Japan Participation Group, Johns Hopkins University, the Joint Institute for 
Nuclear Astrophysics, the Kavli Institute for Particle Astrophysics and 
Cosmology, the Korean Scientist Group, the Chinese Academy of Sciences 
(LAMOST), Los Alamos National Laboratory, the Max-Planck-Institute for 
Astronomy (MPIA), the Max-Planck-Institute for Astrophysics (MPA), New Mexico 
State University, Ohio State University, University of Pittsburgh, University 
of Portsmouth, Princeton University, the United States Naval Observatory, and 
the University of Washington.

\clearpage

\begin{figure}
\plotone{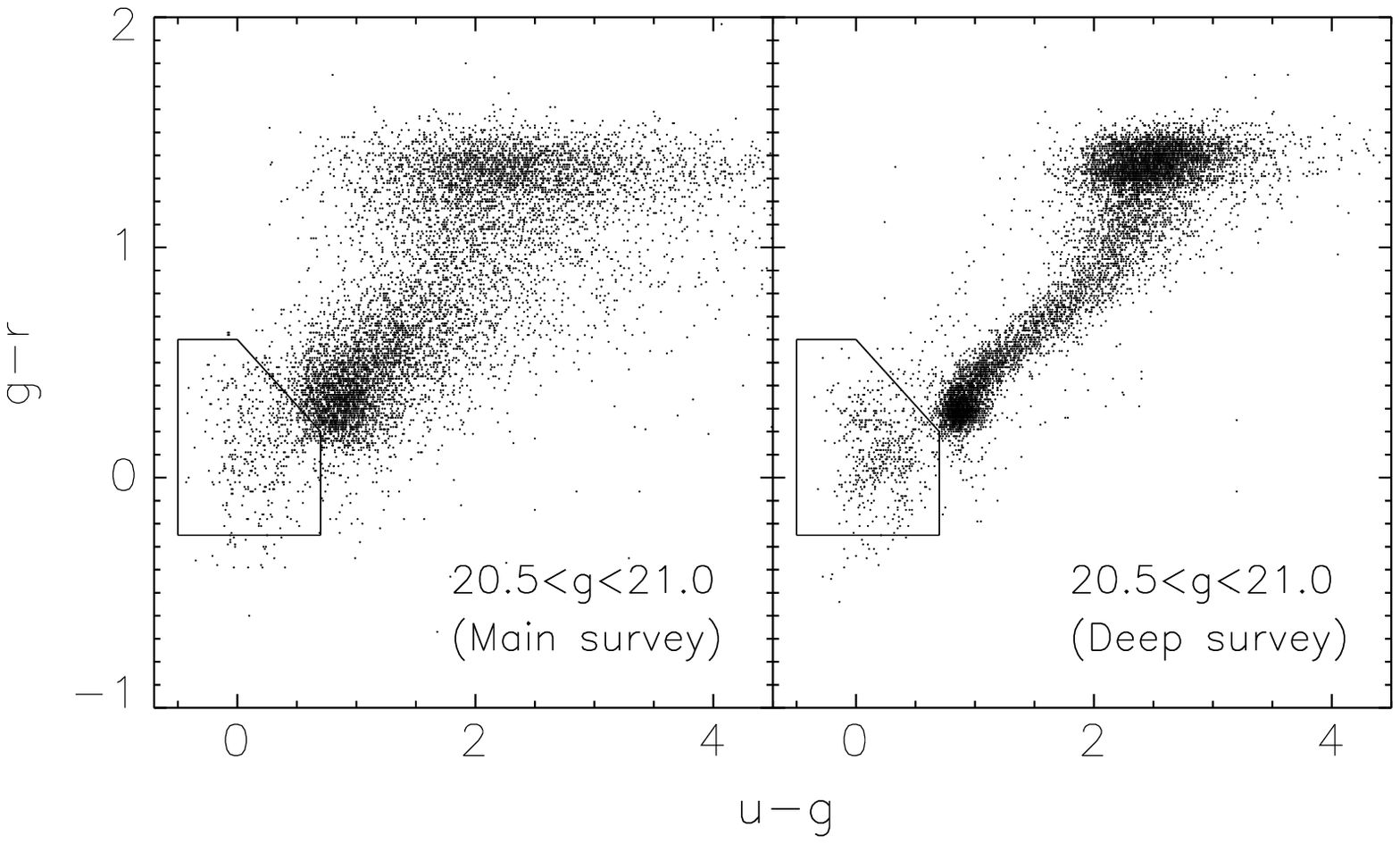}
\caption{
   $u-g$, $g-r$ color-color diagrams for point sources ($20.5<g<21.0$)
   in the SDSS main survey and deep survey with $N_{epoch}\sim13$.
   Each panel includes 10,000 objects. Compared to the main survey, the
   stellar locus in the deep survey is more concentrated, and $UVX$ quasar
   candidates (confined by solid lines) are well separated from the
   stellar locus.
   }
\end{figure}

\clearpage
                                                                                
\begin{figure}
\plotone{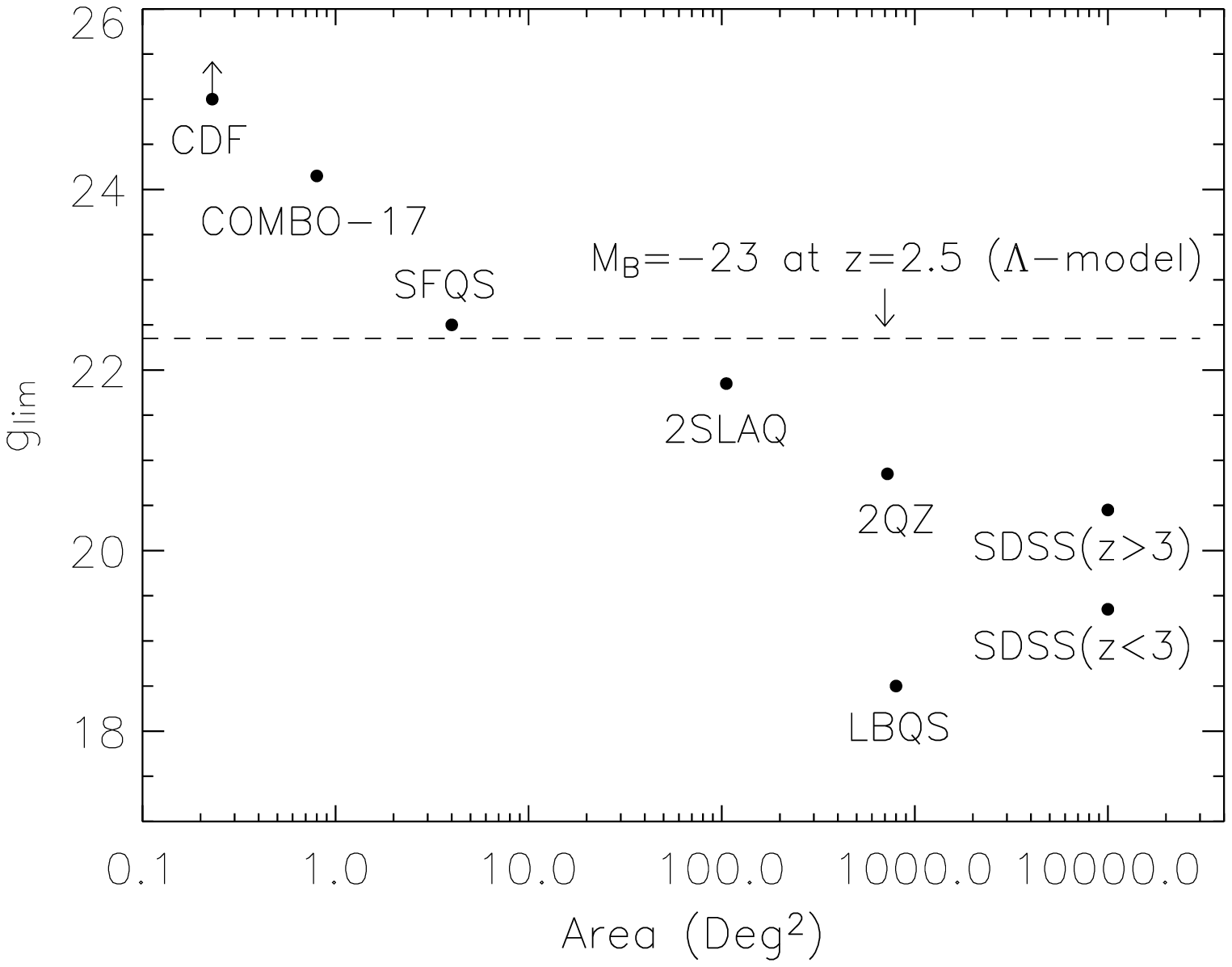}
\caption{
   The comparison between the SFQS and the LBQS \citep{fol87},
   2QZ \citep{boy00,cro04}, SDSS \citep{ric02}, 2SLAQ \citep{ric05},
   COMBO-17 \citep{wol03} and CDF \citep[the area given here is the total 
   area of CDF-N and CDF-S]{bar05} surveys of quasars and AGNs.
   }
\end{figure}
                                                                                
\clearpage

\begin{figure}
\epsscale{.60}
\plotone{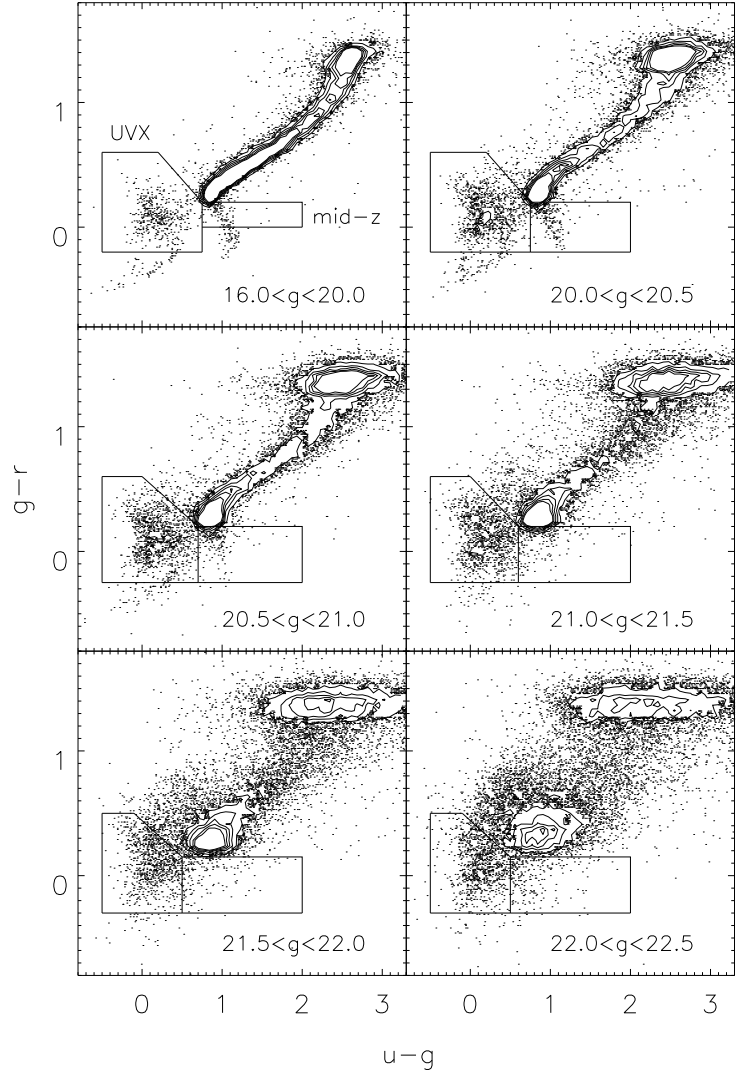}
\caption{
   The selection of $UVX$ and mid-$z$
   candidates from $g=16.0$ to 22.5 in $u-g$,
   $g-r$ diagrams. The inner parts of the diagrams are shown in contours,
   which linearly increase inwards in the density of stars. For the purpose of
   comparison, each panel includes 20,000 objects. The regions
   confined by solid lines are our selection of $UVX$ and mid-$z$ candidates.
   }
\end{figure}

\clearpage

\begin{figure}
\epsscale{1.0}
\plotone{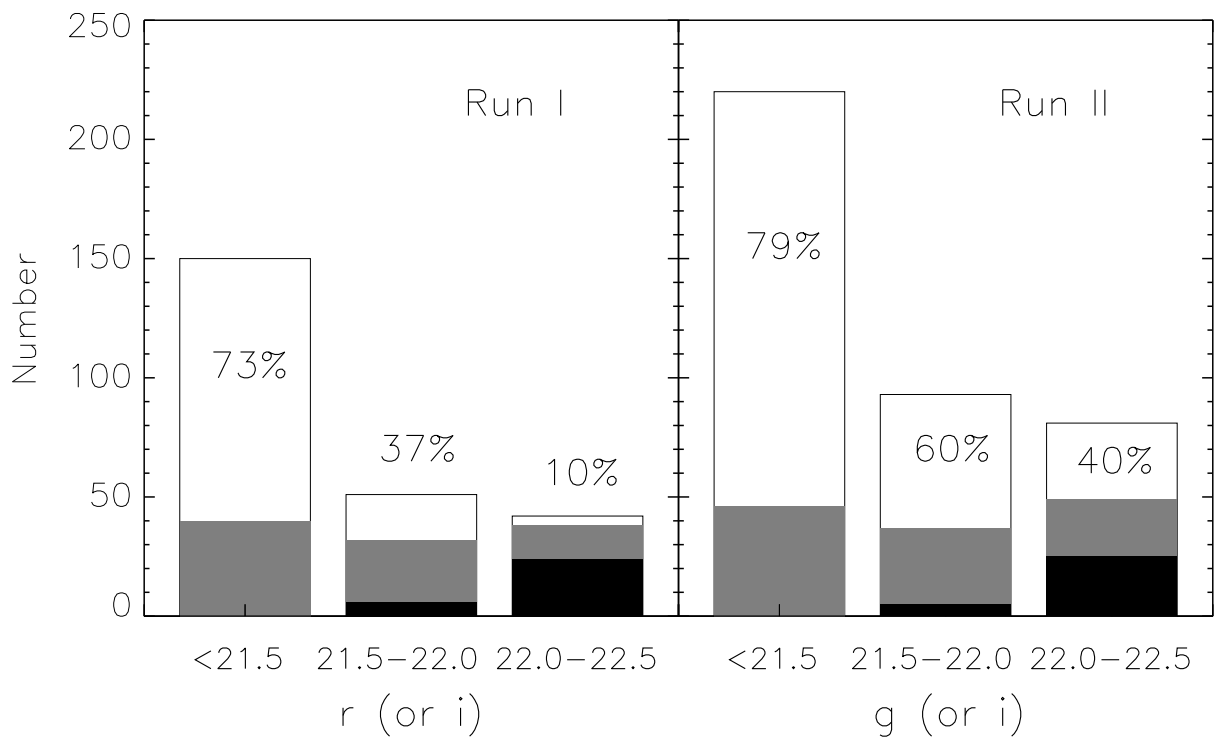}
\caption{
   The numbers of candidates observed and the selection efficiencies in the 
   two runs. The black areas are spectroscopically unidentifiable objects, the 
   gray areas are identified as non-quasars, and the blank regions are 
   identified as quasars. The fractions of quasars are also given within or 
   above the bars. Note that $i$ magnitude is used for high-$z$ ($z\ga 3$) 
   candidates.
   }
\end{figure}

\clearpage

\begin{figure}
\plotone{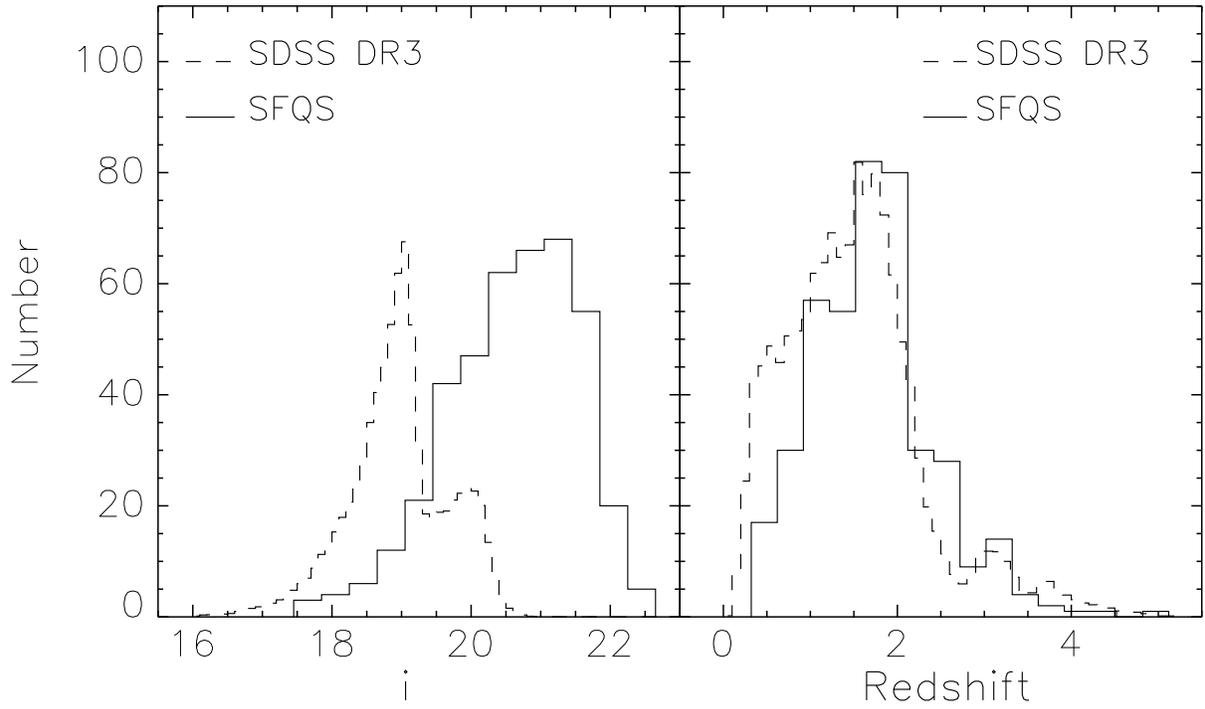}
\caption{
   The $i$ magnitude and redshift distributions of the SFQS
   sample compared with the SDSS main survey. The dashed profiles are from the
   SDSS DR3 \citep{sch05}, and have been scaled to compare with our survey.
   }
\end{figure}
                                                                                
\clearpage

\begin{figure}
\epsscale{0.6}
\plotone{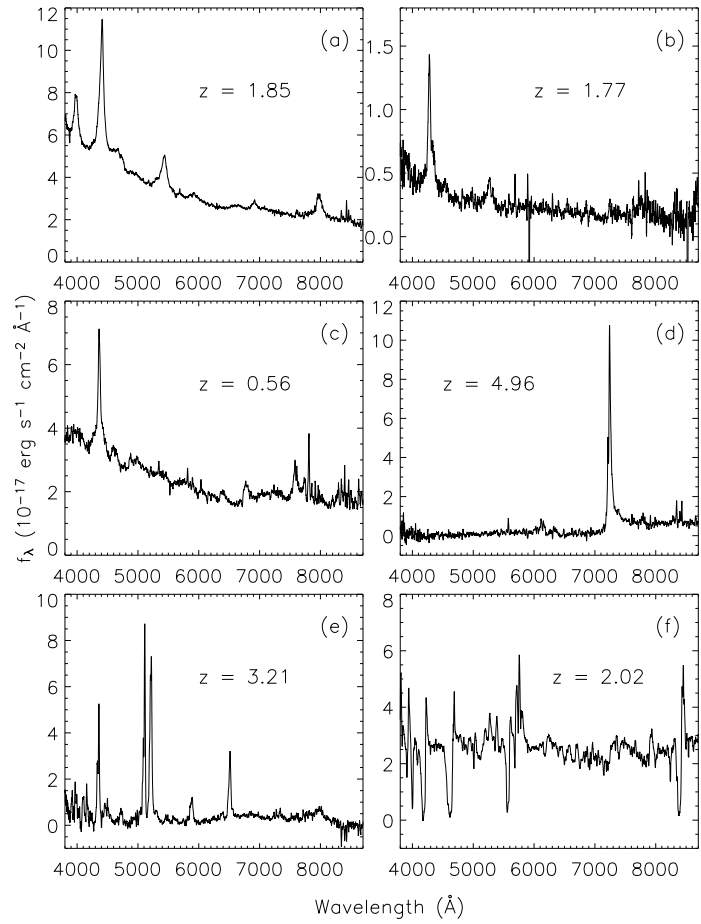}
\caption{
   Six sample spectra obtained by MMT/Hectospec. The spectra are binned
   by 11 pixels. The redshift of each
   quasar is also given in the figure.
   }
\end{figure}
                                                                                
\clearpage

\begin{figure}
\plotone{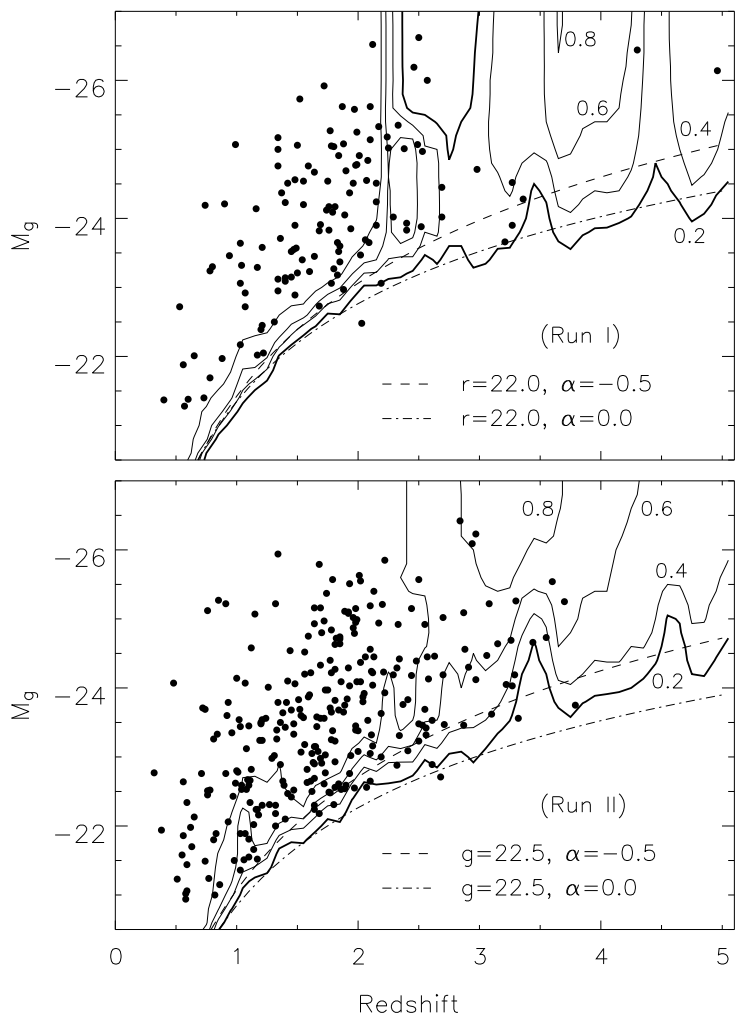}
\caption{
   Selection function of faint quasars as a function of $M_g$ and $z$. The 
   contours are selection probabilities from 0.2 to 0.8 with an interval of 
   0.2. The solid circles are the locations of quasars in our sample.
   Heavy lines (probability $=0.2$) illustrate the limiting magnitudes.
   }
\end{figure}

\clearpage

\begin{figure}
\plotone{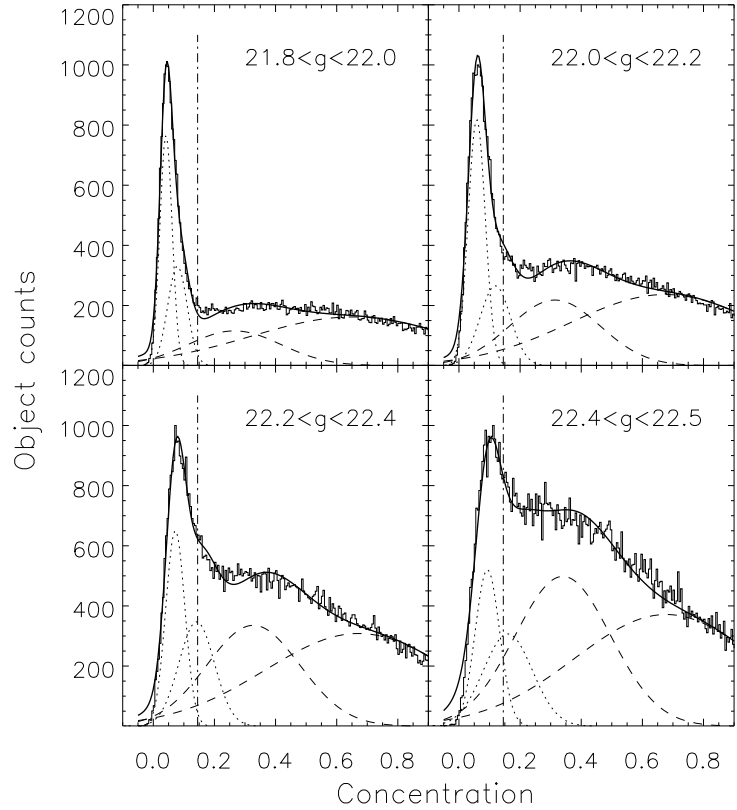}
\caption{
   Correction for the morphology bias.
   The vertical dot-dashed lines separate $stars$ and $galaxies$ by
   the definition of the SDSS photometric pipeline.
   Dotted and dashed lines are the best fits of the double Gaussian 
   components of $stars$ and $galaxies$, respectively. Solid lines are the
   sum of all components.
   }
\end{figure}

\clearpage
                                                                                
\begin{figure}
\plotone{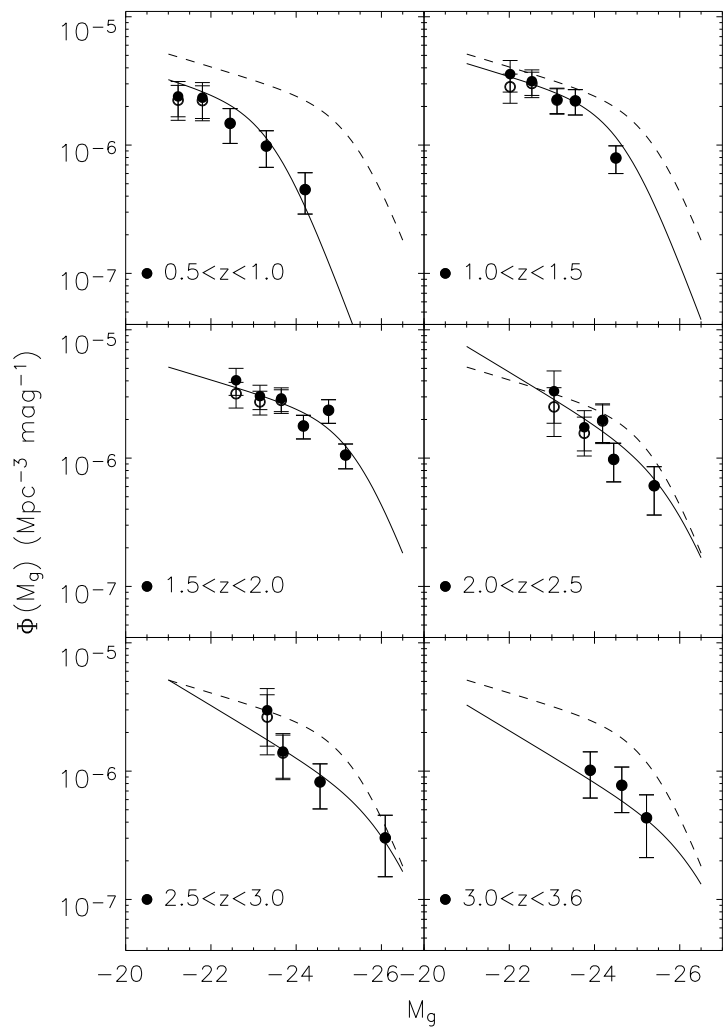}
\caption{
   QLF derived from the traditional $1/V_{a}$ method. Solid
   circles represent the QLF corrected for all four incompletenesses in 
   $\S$ 4.1, while open circles represent the QLF corrected for all 
   incompletenesses except the spectroscopic incompleteness. Solid lines are 
   the best model fits from  $\S$ 4.2. 
   For comparison, the dashed line is the best model fit of the QLF at
   $1.5<z<2.0$.
   }
\end{figure}

\clearpage
                
\begin{figure}
\plotone{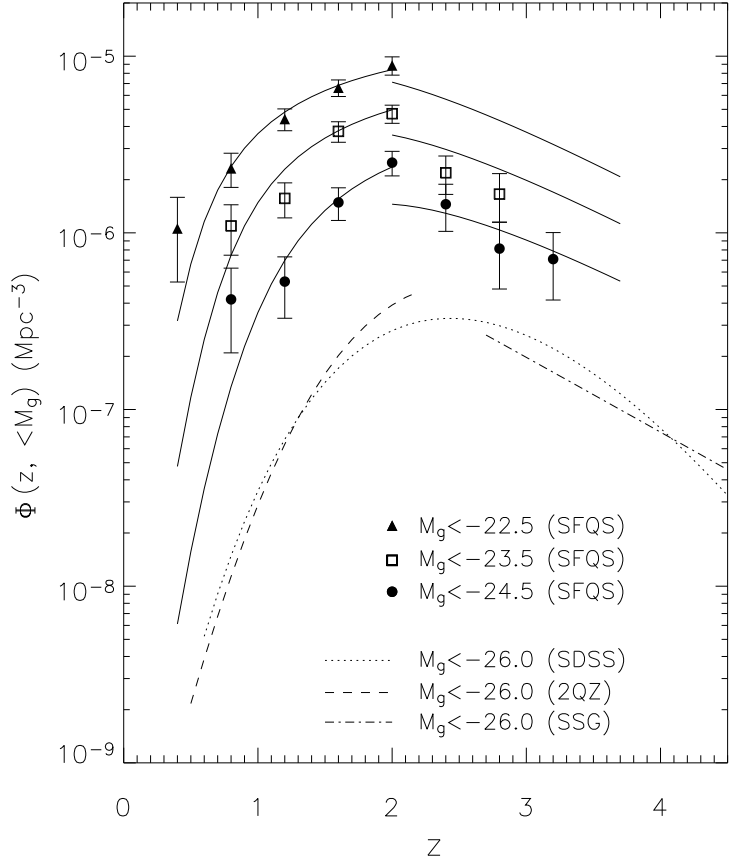}
\caption{
   Integrated comoving density as a function of $z$ for three
   magnitude ranges, $M_g<-22.5$, $-23.5$, and $-24.5$, respectively. The
   redshift bins are from $z=0.4$ to 3.6 with the interval of $\Delta z=0.4$.
   We do not include incomplete bins.
   Solid curves are integrated densities calculated from the best
   model fits in $\S$ 4.2, while dotted, dashed and dot-dashed curves are the
   integrated densities from the SDSS \citep{ric06}, 2QZ \citep{boy00},
   and SSG \citep{sch95}, respectively. 
   }
\end{figure}
                                                                                
\clearpage
                                                                
\begin{figure}
\plotone{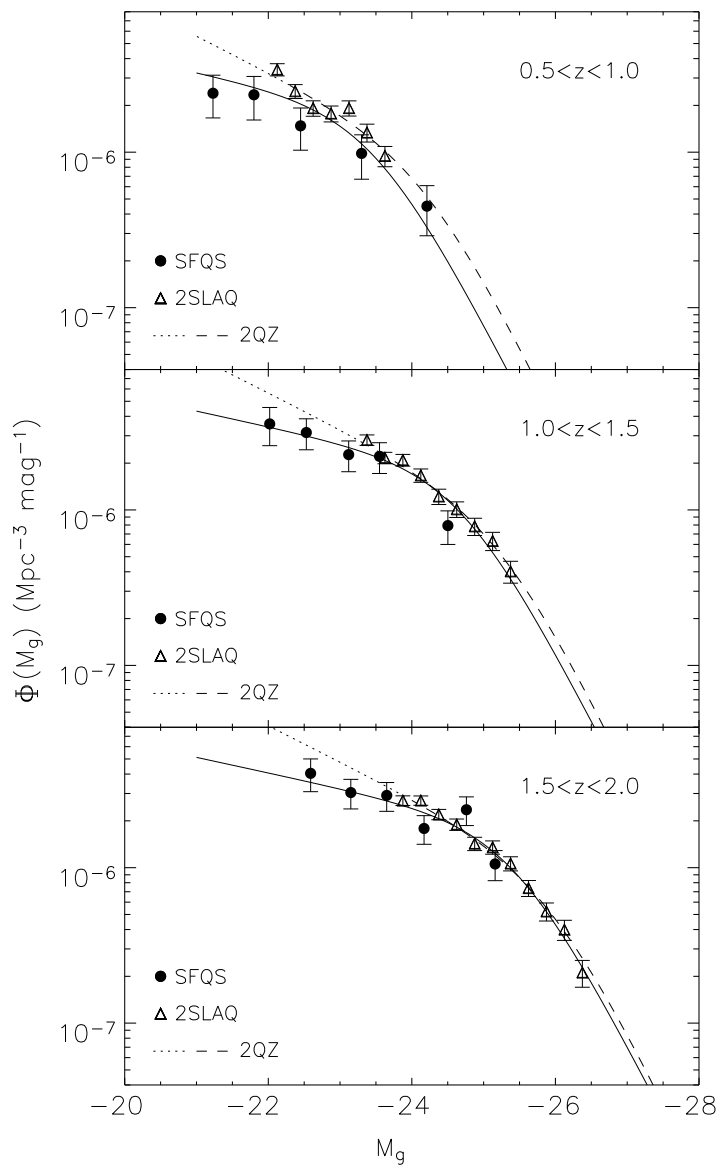}
\caption{
   Comparison with the 2QZ \citep{boy00} and 2SLAQ \citep{ric05}. Solid 
   circles and open triangles are the SFQS QLF and 2SLAQ QLF, respectively.
   The dotted and dashed lines represent the 2QZ QLF, calculated from the 
   best model fits of \citet{boy00}. The dashed-line parts are roughly the 
   range of luminosity that the observed 2QZ QLF actually covered 
   \citep{boy00}. The solid lines are the best model fits of the SFQS QLF.
   }
\end{figure}

\clearpage

\begin{figure}
\plotone{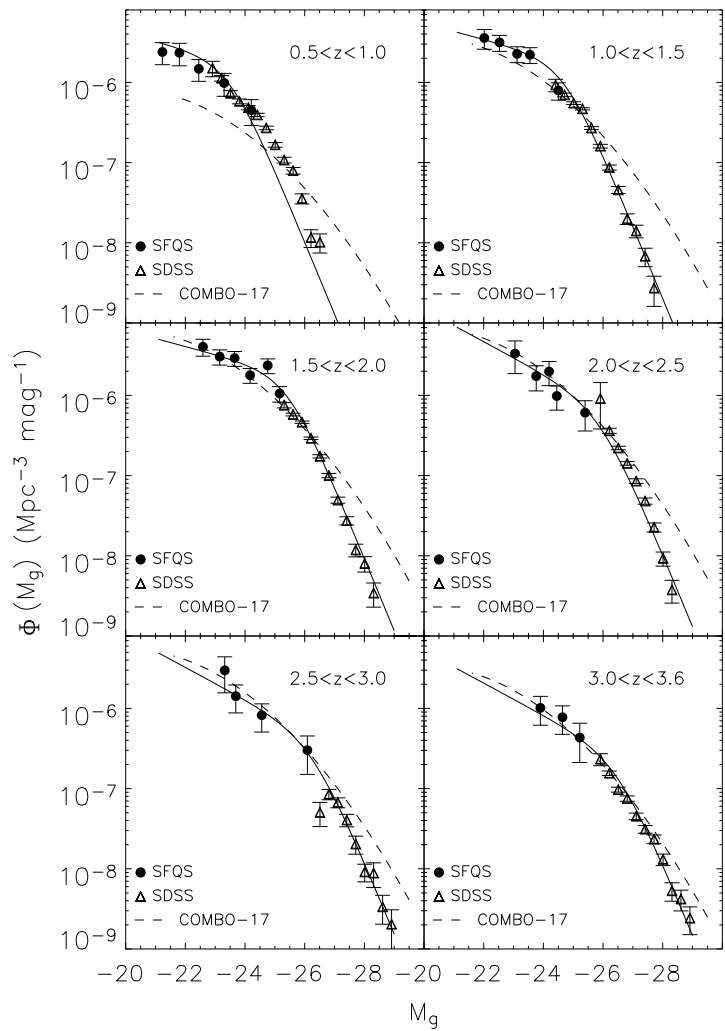}
\caption{
   Comparison with the SDSS \citep{ric06} and COMBO-17 \citep{wol03}.
   Solid circles and open triangles are the SFQS QLF and SDSS QLF, 
   respectively. The dashed lines represent the COMBO-17 QLF, calculated
   from the best model fits of \citet{wol03}. The solid lines are the
   best model fits of the SFQS QLF.
   }
\end{figure}

\clearpage

\begin{figure}
\plotone{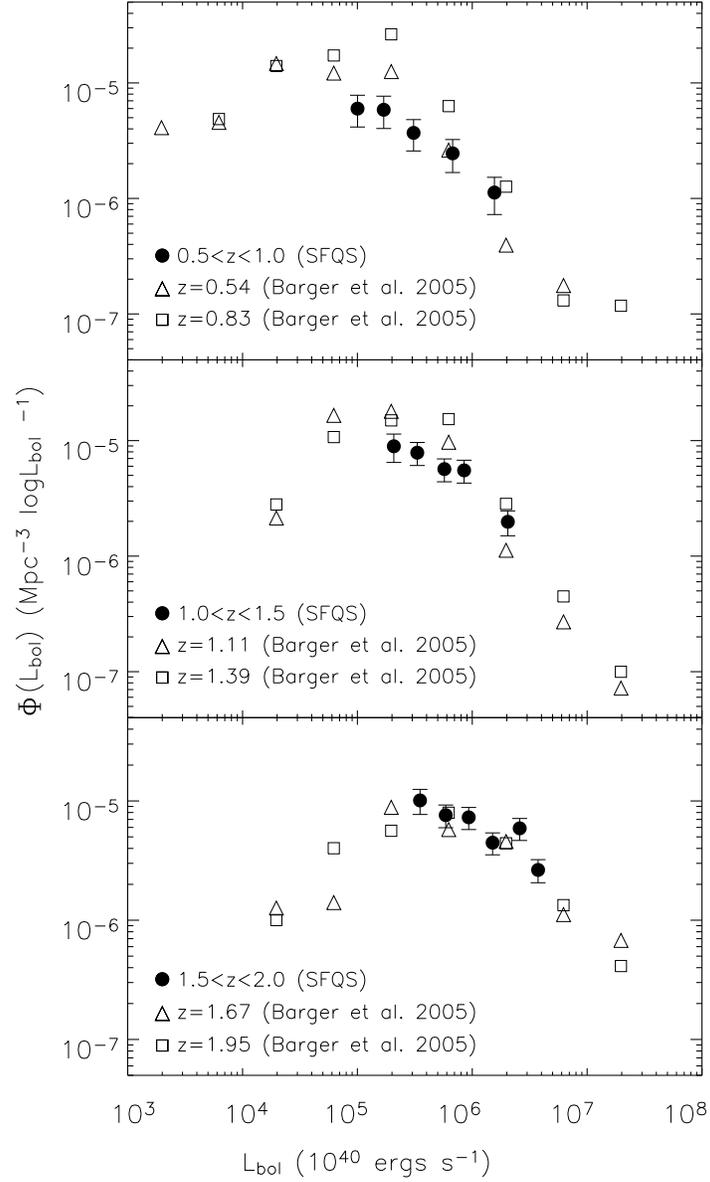}
\caption{
   Comparison with the CDF survey \citep{bar05}. Solid circles represent
   the SFQS QLF. Open triangles and squares are from Figure 22 of 
   \citet{bar05}.
   }
\end{figure}

\clearpage
\begin{table}
\caption{Central positions and exposure time for the 5 Hectospec fields}
\begin{tabular}{ccccc}
\tableline\tableline
 & Date\tablenotemark{a} & RA(J2000) & Dec(J2000) & $t_{exp} $(min) \\
\tableline
Run I  &Jun 13&$21^{h}34^{m}00{\fs}0$&$00{\degr}06{\arcmin}00{\farcs}0$&100\\
       &Jun 15&$21^{h}34^{m}00{\fs}0$&$00{\degr}06{\arcmin}00{\farcs}0$&100\\
       &Jun 20&$21^{h}30^{m}00{\fs}0$&$00{\degr}06{\arcmin}00{\farcs}0$&100\\
Run II &Nov 09&$23^{h}16^{m}00{\fs}0$&$-00{\degr}06{\arcmin}00{\farcs}0$&120\\
       &Nov 11&$23^{h}40^{m}00{\fs}0$&$-00{\degr}06{\arcmin}00{\farcs}0$&180\\
       &Nov 13&$02^{h}04^{m}00{\fs}0$&$-00{\degr}06{\arcmin}00{\farcs}0$&80\\
       &Nov 19&$02^{h}04^{m}00{\fs}0$&$-00{\degr}06{\arcmin}00{\farcs}0$&120\\
\tableline
\end{tabular}
\tablenotetext{a}{Dates in 2004}
\end{table}

\clearpage
\begin{table}
\caption{Quasar sample for the SFQS survey}
\begin{tabular}{cccccc}
\tableline\tableline
Name (J2000 Coordinates) & Redshift & $g$ & $M_{g}$ & $\alpha$\\
\tableline
SDSS J231601.68$-$001237.0  &  2.00  &  21.69  &  -23.08  &  -0.23  \\
SDSS J231548.40$-$003022.7  &  1.53  &  20.73  &  -23.60  &  -0.44  \\
SDSS J231527.54$-$001353.8  &  1.33  &  20.97  &  -23.26  &  -0.54  \\
SDSS J231541.51$-$003137.2  &  1.72  &  21.73  &  -22.61  &  -0.31  \\
SDSS J231422.26$+$000315.7  &  3.22  &  21.38  &  -24.05  &   0.10  \\
SDSS J231442.27$-$000937.2  &  3.30  &  21.09  &  -25.26  &  -0.49  \\
SDSS J231534.22$-$002610.0  &  0.58  &  21.29  &  -21.02  &  -1.07  \\
SDSS J231519.33$-$001129.5  &  3.06  &  22.17  &  -24.47  &  -1.10  \\
SDSS J231504.04$-$001434.2  &  2.10  &  22.17  &  -23.05  &  -0.41  \\
SDSS J231446.30$-$002206.9  &  2.97  &  21.97  &  -24.12  &  -0.69  \\
$\cdot \cdot \cdot $ &        &         &          &         \\
\tableline
\end{tabular}
\tablecomments{Table 2 is given in the electronic edition. The typical errors
 for a $g$ = 21.0 quasar at $z$ = 2.0 are, $\sigma_{z}<$ 0.01, $\sigma_{M_g}$
 = 0.08, and $\sigma_{\alpha}$ = 0.15.}
\end{table}

\end{document}